\documentclass[reprint,amsmath,nofootinbib,amssymb,aps,showpacs,prc]{revtex4-1}
\usepackage{graphicx}
\usepackage{subfigure}
\usepackage{multirow}
\usepackage{mathrsfs}
\usepackage[colorlinks,
linkcolor=blue,
anchorcolor=blue,
citecolor=blue]{hyperref}
\usepackage[all]{hypcap}
\usepackage{epstopdf}
\usepackage{caption}
\usepackage{color}
\usepackage{float}
\usepackage{bm}

\begin{document}

%%%%%%%%%%%%%%%%%%%%%%%%%%%%%%%%%%%%%%%%%%%%%%%%%%%%%%%%%
%%%%%%%%%%%%%%%%%%%%%%%%%%%%%%%%%%%%%%%%%%%%%%%%%%%%%%%%%
%%%%%%%%%%%%%%%%%%%%%%%%%%%%%%%%%%%%%%%%%%%%%%%%%%%%%%%%%
\title{Effect of magnetic fields on pairs of oppositely charged 
particles in ultrarelativistic heavy-ion collisions}
%\collaboration{the STAR Collaboration}
%\date{\today}

\author{Y. J. Ye$^{1,2,3}$, Y. G. Ma$^{1,2}$, A. H. Tang$^4$ and G. Wang$^5$}
\affiliation{
\mbox{$^1$ Key Laboratory of Nuclear Physics and Ion-beam Application (MOE),} \\
\mbox{Institute of Modern Physics, Fudan University, Shanghai 200433, P.R. China}
\mbox{$^2$Shanghai Institute of Applied Physics, Chinese Academy of Sciences, Shanghai 201800, P.R. China}\\
\mbox{$^3$ University of Chinese Academy of Sciences, Beijing 100049, P.R. China}
\mbox{$^4$Brookhaven National Laboratory, Upton, New York 11973, USA}\\
\mbox{$^5$ Department of Physics and Astronomy, University of California, Los Angeles, California 90095, USA}
}

\begin{abstract}
The initial strong magnetic field produced in high-energy nuclear collisions will distort the distribution of the relative angle between oppositely charged particles within a pair. In this paper, two experimental observables are examined to quantify such effects: one based on the framework for detecting the hyperon global polarization, and the other based on the balance function. We also discuss the optimization of the signal, as well as the expected magnitude ranges for the two observables.
\end{abstract}


\maketitle

%%%%%%%%%%%%%%%%%%%%%%%%%%%%%%%%%%%%%%%%%%%%%%%%%%%%%%%%%
%%%%%%%%%%%%%%%%%%%%%%%%%%%%%%%%%%%%%%%%%%%%%%%%%%%%%%%%%
%%%%%%%%%%%%%%%%%%%%%%%%%%%%%%%%%%%%%%%%%%%%%%%%%%%%%%%%%
\section{Introduction}\label{sec:intro}
In noncentral ultra-relativistic heavy-ion collisions, spectator protons pass by each other at nearly the speed of light, producing ultra-strong magnetic fields~\cite{Kharzeev:2007jp}. Such an enormous magnetic field has many interesting consequences. For example, when acting together with quantum anomalies, the magnetic field can enhance the anisotropy of the soft-photon production~\cite{Basar:2012bp}, and polarize photons in opposite ways in the hemispheres above and below the reaction plane~\cite{Hirono:2015rla,Ipp:2007ng,Mamo:2013jda}. In particular, when a local domain in the collision system obtains a non-zero topological charge, the interplay between the strong magnetic field and this topological charge can induce electric charge separation with respect to the reaction plane --- the chiral magnetic effect (CME)~\cite{Kharzeev:2007jp,Kharzeev:2015znc,Huang}. This phenomenon, if confirmed, would indicate the local parity violation in the strong interaction. Similarly, under the same strong magnetic field, a non-zero chemical potential of electric charge can also lead to chiral charge separation, the chiral separation effect (CSE). Furthermore, the CME and the CSE can even feed each other, forming a chiral magnetic wave (CMW)~\cite{Burnier:2011bf}. With these important implications on the fundamental property of the QCD vacuum, both the CME and the CMW have been intensively studied at RHIC and the LHC~\cite{Abelev:2009ac,Abelev:2009ad,Adamczyk:2013hsi,Adamczyk:2014mzf,Adamczyk:2013kcb,Adamczyk:2015eqo,Acharya:2017fau,Adam:2015vje,Abelev:2012pa,Khachatryan:2016got,Sirunyan:2017quh}, though the observation of these effects in heavy-ion collisions is not yet conclusive. See~\cite{Kharzeev:2015znc} for a progress review. 

One of the preconditions for the aforementioned novel phenomena is the initial ultra-strong magnetic field, which has not been directly detected. Several probes of the magnetic field have been proposed, such as the anisotropic charmonium production~\cite{Guo:2015nsa}, the separation in the global polarization between $\Lambda$ and $\bar{\Lambda}$~\cite{STAR:2017ckg}, the directed flow of charm quarks~\cite{Das:2016cwd}, and recently, the $p_{T}$ broadening of $e^+e^-$ spectra~\cite{Adam:2018tdm}. In this paper, we inspect the imprint left by the magnetic field on pairs of particles with opposite charges. At midrapidities, the motion of charged particles could be affected by two competing mechanisms, namely the Lorentz force and the Faraday's law effect~\cite{Gursoy:2014aka,Gursoy:2018yai}: the former is caused by the initial strong magnetic field, and the latter, by the fast decline of the field. This paper will focus on the net effect of the two. We have no intention to (and indeed by this study we cannot) separate the two effects.

\section{Distortion of the relative angle between electron and positron within a pair}\label{sec:distortion}
\vspace{-0.08cm}

\begin{figure}[htbp]
\centering
\makebox[1cm]{\includegraphics[width=0.45 \textwidth]{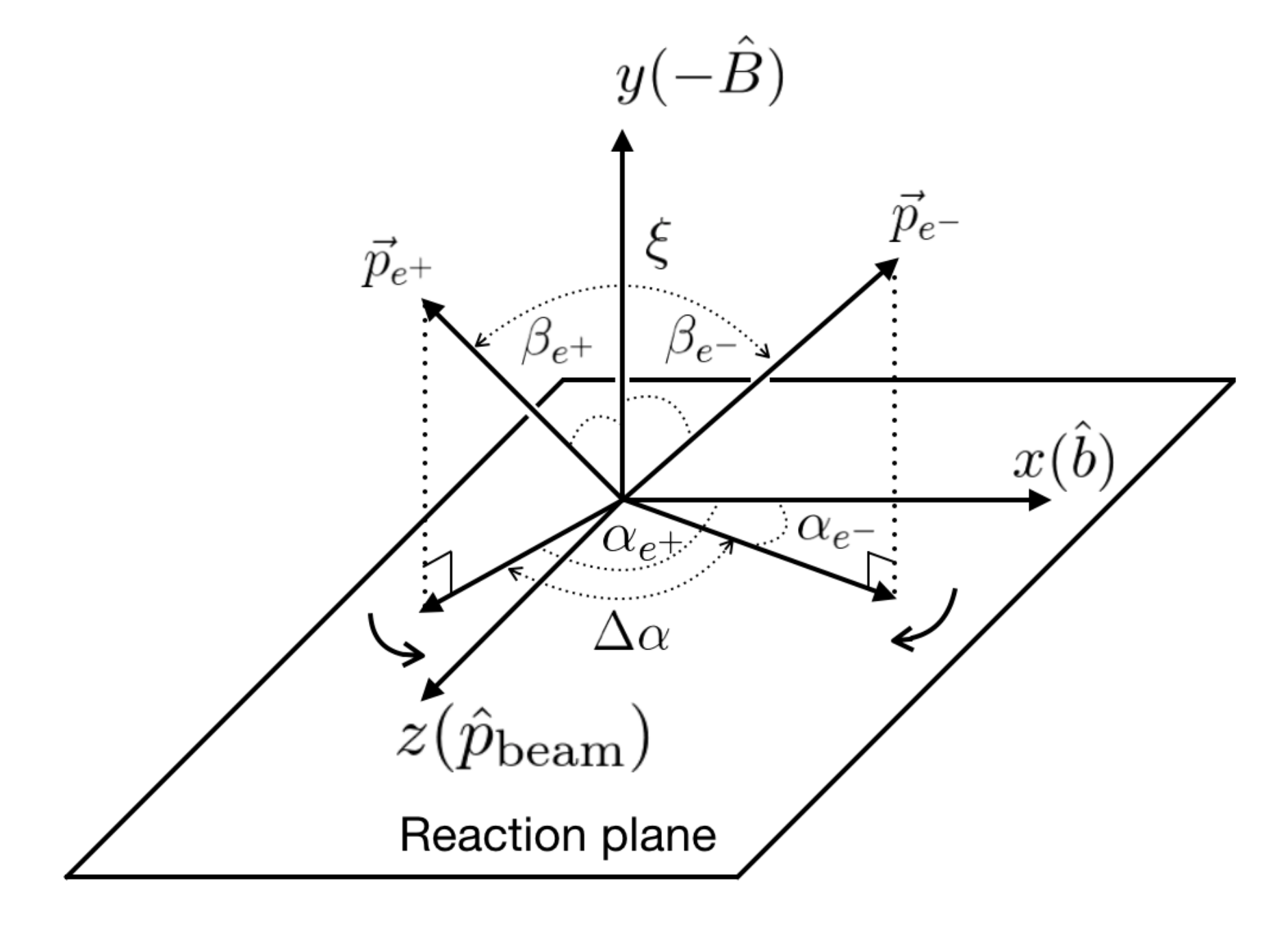}}
\caption{The setup of the coordinate system. See text for details.}
\label{fig:schematic}
\end{figure}

The coordinate system in this study is delineated in Fig.~\ref{fig:schematic}, the same as that in Ref.~\cite{Abelev:2007zk}. The $x$-axis is set by the direction of the impact parameter ($\hat{b}$), and the $z$-axis represents the beam direction ($\hat{p}_{\mathrm{beam}}$).
The $y$-axis points to the opposite direction of the magnetic field  ($\hat{B}=\hat{b} \times \hat{p}_\mathrm{beam}$). 

\begin{figure}[htbp]
\centering
\makebox[1cm]{\includegraphics[width=0.45 \textwidth]{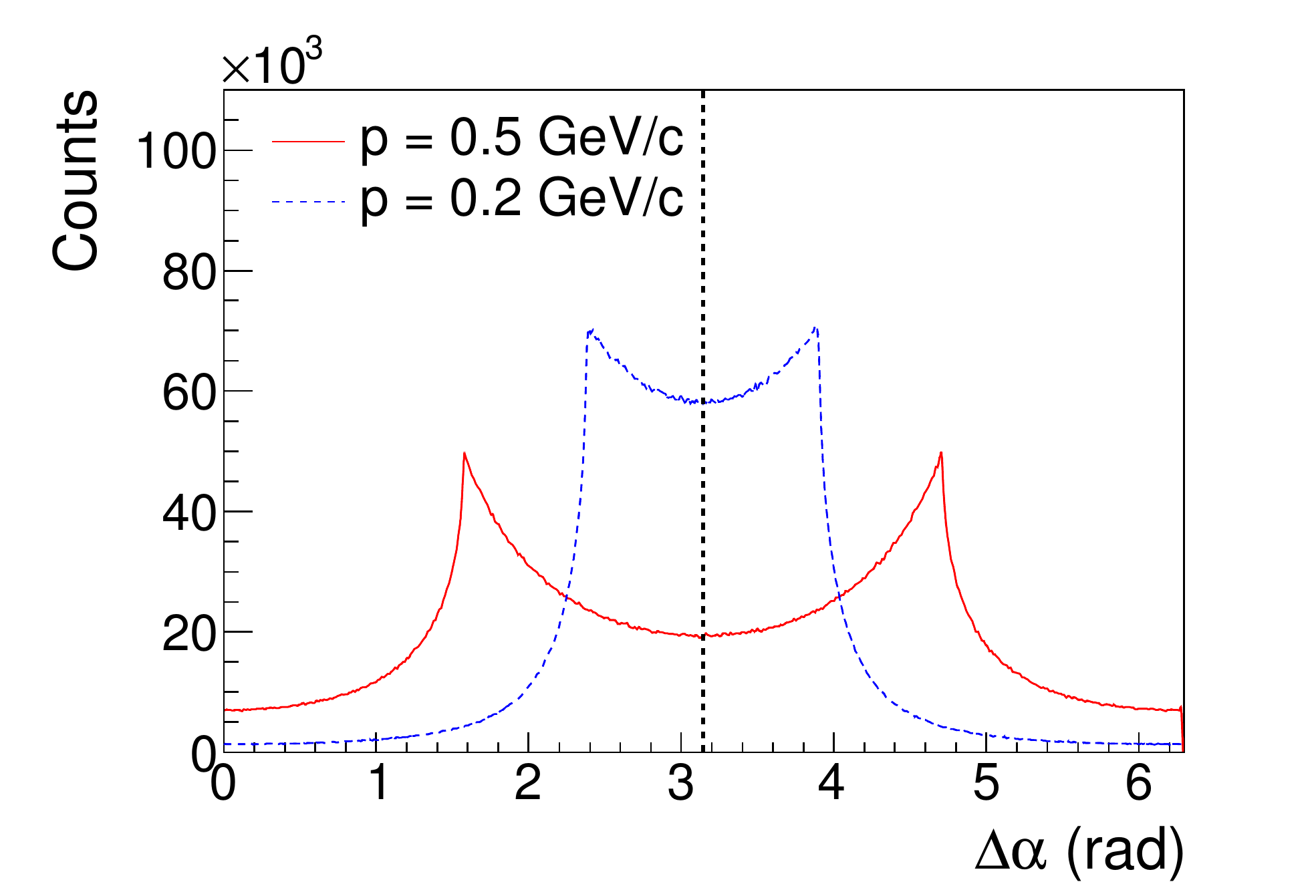}}
\caption{(Color online) Simulated $\Delta\alpha$ distributions for $e^+e^-$ pairs from decays of virtual particles with mass of 0.5 GeV/$c^2$, and momentum of 0.5 GeV/$c$ (solid line) or 0.2 GeV/$c$ (dotted line). The momentum directions of the virtual particles have been randomized.}
\label{fig:structure_move}
\end{figure}

Without loss of generality, we assume that the Lorentz force dominates over the Faraday's law effect at midrapidities. As a result, looking down the $y$-axis,  negatively- and positively-charged particles are deflected clockwise and counter-clockwise, respectively  (Fig.~\ref{fig:schematic}). Such deflections are most prominent for electrons and positions owing to their light masses. Another reason to take leptons as an example is that they are ``penetrating probes", receiving minimal influence from later-stage hadronic interactions. We introduce a relative angle 
\begin{eqnarray}
\Delta\alpha\equiv\alpha_{e^{+}}-\alpha_{e^{-}},
\label{eq:alpha}
\end{eqnarray}
where $\alpha_{e^{-}}$ ($\alpha_{e^{+}}$)  is the angle of electron(positron) momentum projected onto the reaction plane (Fig.~\ref{fig:schematic}). If charged particles are produced randomly, they remain random after the deflections. In this case, the $\Delta\alpha$ distribution is flat before and after the deflections, and one cannot tell whether particles have experienced deflections or not. In reality, particles are not always produced randomly. Oppositely charged particles that originate from real particle decays are governed by energy and momentum conservation, causing a non-uniform structure in the $\Delta\alpha$ distribution. Figure~\ref{fig:structure_move} illustrates the $\Delta\alpha$ distribution for positron-electron pairs that come from decays of virtual particles with mass of 0.5 GeV/$c^2$, and momentum of 0.2 GeV/$c$ or 0.5 GeV/$c$. The momentum directions of the virtual particles have been randomized. By construction, without a magnetic field, the $\Delta\alpha$ distribution is symmetric w.r.t $\Delta \alpha = \pi$, since the probability has to be the same when the two decay daughters are swapped. The opening angle between the two daughters depends on the parent's mass and momentum: larger mass and/or lower momentum will lead to a wider opening angle. In Fig.~\ref{fig:structure_move} this means that the two peak positions will move towards $\Delta \alpha = \pi $ for larger mass and/or lower momentum. In reality, it is not straightforward to predict the  structure of this distribution, which is a convoluted effect of mass and momentum of the $e^+e^-$ continuum, but in general, the distribution is non-flat and symmetric w.r.t $\Delta \alpha = \pi$.

-eps-converted-to.pdf

In the presence of the magnetic field, charged particles at midrapidities are deflected by both the Lorentz force and the Faraday's law effect. Below we take the Lorentz force as an example to describe our proposed observable. The Lorentz force will cause an angle change, $\delta\alpha$, to a particle with charge $q$ and mass $m$,
\begin{eqnarray}
\delta\alpha = - \frac{q \int B dt}{\gamma_{{\rm RP}} m},
\label{eq:e_shift}
\end{eqnarray}
where $t$ is the acting time of the magnetic field ($B$), and $\gamma_{{\rm RP}}$ is the Lorentz factor for the particle's velocity projected onto the reaction plane. The change in $\Delta \alpha(\equiv \alpha_{e^+} - \alpha_{e^-})$ for an $e^+e^-$ pair is 
\begin{eqnarray}
\begin{aligned}
\delta (\Delta \alpha) &= \delta\alpha_{e^+} - \delta\alpha_{e^-} \\
&= - \frac{|e|\int B dt}{m_e} ( \frac{1}{\gamma_{\rm RP,e^+}} + \frac{1}{\gamma_{\rm RP,e^-} } ).
\end{aligned}
\label{eq:e_deltaAlphaChange}
\end{eqnarray}
\begin{figure}[htbp]
\centering
\makebox[1cm]{\includegraphics[width=0.45 \textwidth]{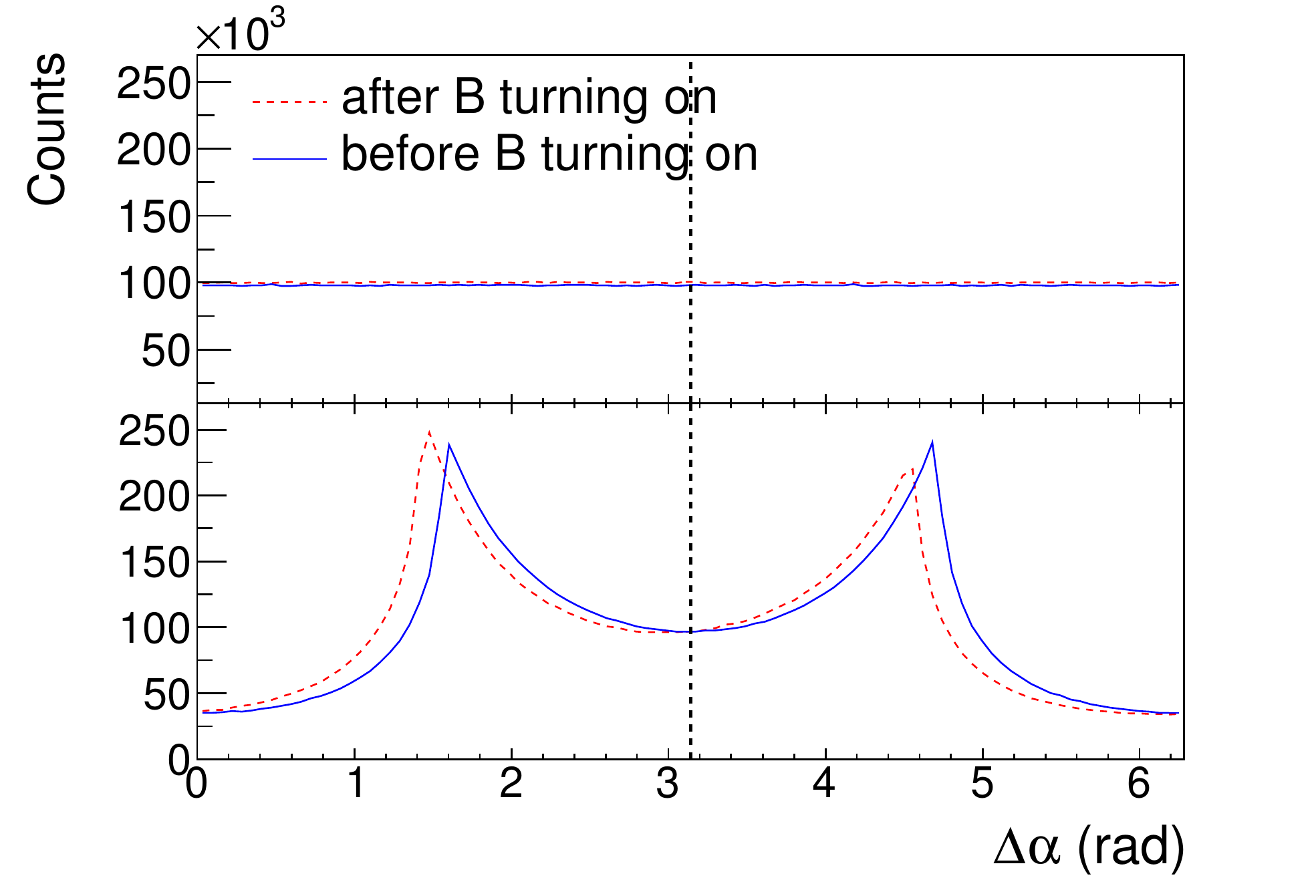}}
\caption{(Color online) The upper panel shows the $\Delta\alpha$ distribution for random electrons and positrons, before (solid line) and after (dashed line) the turning on of the magnetic field. The lower panel shows the same distribution for $e^+e^-$ pairs originating from decays of virtual particles with mass of 0.5 GeV/$c^2$ and momentum of 0.5 GeV/$c$, before (solid line) and after (dashed line) the turning on of the magnetic field. }
\label{fig:compare_rand_sig}
\end{figure}

Positively- and negatively-charged particles receive the Lorentz force in opposite directions, and the deflection of an electron or a positron depends on its own $\gamma_{{\rm RP}}$ (Eq.(\ref{eq:e_deltaAlphaChange})). Consequently, the $\Delta \alpha$ distribution is distorted, with its peak shifted and skewed. To demonstrate this effect, we apply a magnetic field of $e B = 5 \times 10^{-3}  \mathrm{\, GeV^2} $ for a period of 1 fm/$c$ to one of the aforementioned simulation cases ($p = 0.5$ GeV/$c$), as shown in the lower panel of Fig.~\ref{fig:compare_rand_sig}. The introduction of the magnetic field clearly breaks the symmetry w.r.t $\Delta \alpha = \pi$. In comparison, a case with random electrons and positrons is shown in the upper panel where no effects can be seen.

\begin{figure}[htbp]
\centering
\makebox[1cm]{\includegraphics[width=0.45 \textwidth]{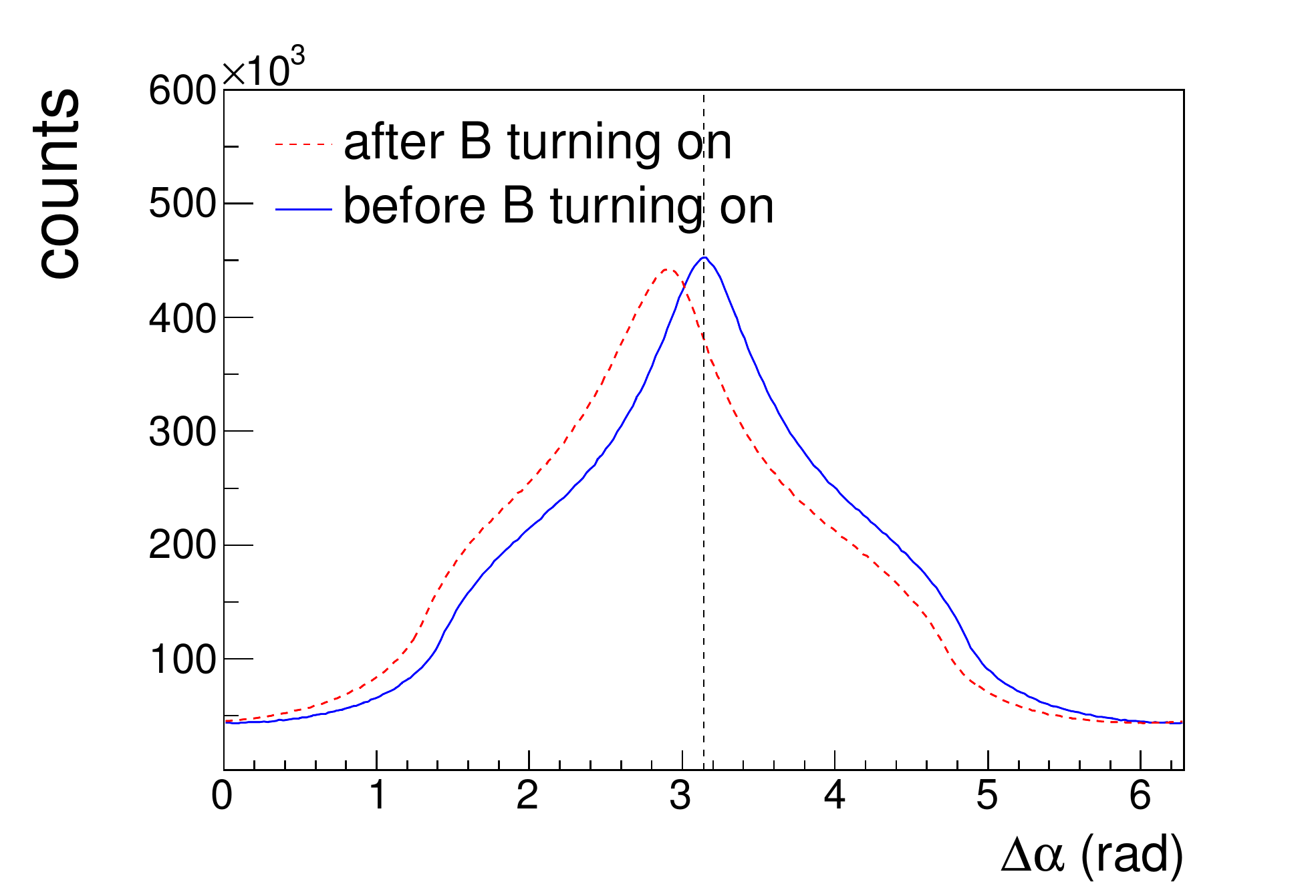}}
\caption{(Color online) The simulated $\Delta\alpha$ distribution using the $e^+e^-$ continuum taken from the STAR publication~\cite{Adam:2018tdm}, in which the $e^+e^-$ continuum is for $p_T < 1.5$ GeV/$c$, $|\eta| < 1$ and invariant mass from 0.4 GeV/$c^2$ to 0.76 GeV/$c^2$.  The blue solid line represents the distribution without the magnetic field, and the red dashed line denotes the case with the magnetic field.}
\label{fig:star}
\end{figure}

As an example of the $\Delta \alpha$ distribution with a convoluted mass continuum, we repeat in Fig.~\ref{fig:star} the same procedure but with a realistic $e^+e^-$ continuum taken from the STAR publication~\cite{Adam:2018tdm}. Again the distribution is skewed when the magnetic field is on. The inclusion of the Faraday's law effect would make this distortion less prominent, but in general won't exactly cancel the Lorentz effect. The description of the combination of the two effects (the Lorentz effect and the Faraday's law effect) requires detailed knowledge of the magnetic field and its time dependence, which is beyond the scope of this paper. 
To summarize this section, from the experiment's perspective, the asymmetric $\Delta \alpha$ distribution is a key signature of the initial magnetic field. In the following two sections, we propose two approaches to quantify this effect.

\section{Angular correlation between the $e^+e^-$ pair and the reaction plane%\texorpdfstring{$\langle \mathrm{sin} \Delta \alpha \rangle$}{sinDeltaAlpha}
}\label{sec:SDotB}
\vspace{-0.08cm}

Experimentally the direction of the magnetic field is determined by $\hat{B} = \hat{b} \times \hat{p}_\mathrm{beam}$, and the azimuthal angle of $\hat{b}$ in the transverse plane ($x {\text -} y$ plane in Fig.~\ref{fig:schematic}) is called the reaction plane angle,  $\Psi_{\rm RP}$. $\Psi_{\rm RP}$ itself is reconstructed from final particles, and bears a finite resolution.
%, as event by event the reconstructed event plane is centered around the true reaction plane but not as the same. 
Thus the observed $\Delta \alpha$ is different from the true value, and the observed asymmetry of the $\Delta \alpha$ distribution is also biased. It is not facile to correct the $\Delta \alpha$ distribution for this resolution effect. However, we will show below that with a well-defined observable that is based on $\Delta \alpha$, the effect of the finite event plane resolution can be taken into account to quantify the $\Delta \alpha$ asymmetry.  

We first define a unit vector for an $e^+e^-$ pair
\begin{eqnarray}
\hat{s}\equiv \frac{\hat{e}^+\times\hat{e}^-}{| \sin \xi |},
\label{eq:spin}
\end{eqnarray}
where $\hat{e}^+ (\sin \beta_{\scriptscriptstyle e^+} \cos \alpha_{\scriptscriptstyle e^+}, \cos\beta_{\scriptscriptstyle e^+}, \sin \beta_{\scriptscriptstyle e^+} \sin \alpha_{\scriptscriptstyle e^+})$ and $\hat{e}^- (\sin \beta_{\scriptscriptstyle e^-} \cos \alpha_{\scriptscriptstyle e^-}, \cos \beta_{\scriptscriptstyle e^-}, \sin \beta_{\scriptscriptstyle e^-} \sin \alpha_{\scriptscriptstyle e^-})$ are unit vectors representing the momentum directions of positron and electron inside a pair, respectively, and $\xi$ is the opening angle between them. Let $\theta$ be the angle between $\hat{s}$ and $\hat{B}$, and we have
\begin{eqnarray}
\begin{aligned}
\cos \theta  = \hat{s} \cdot \hat{B} 
= -\frac{ \sin \beta_{\scriptscriptstyle e^+} \sin \beta_{\scriptscriptstyle e^-} \sin \Delta \alpha }{| \sin \xi |}.
\end{aligned}
\label{eq:sdotb}
\end{eqnarray}
Note that $\cos \theta$ is a function of $\beta_{\scriptscriptstyle e^+}$, $\beta_{\scriptscriptstyle e^-}$ and $\Delta \alpha$ only, since $\xi$ is determined by these three angles
\begin{eqnarray}
\cos\xi = \cos\beta_{\scriptscriptstyle e^+} \cos\beta_{\scriptscriptstyle e^-} + \sin\beta_{\scriptscriptstyle e^+}\sin\beta_{\scriptscriptstyle e^-}\cos\Delta\alpha.
\end{eqnarray}
When the magnetic field is turned on, $\Delta \alpha$ decreases, whereas $\beta_{\scriptscriptstyle e^+}$ and $\beta_{\scriptscriptstyle e^-}$ remain unchanged. The change in the $\cos \theta$ distribution will come solely from the change  in $\Delta \alpha$. Thus instead of directly looking at the $\Delta \alpha$ distribution, we can check the asymmetry (w.r.t. $\theta = \pi/2$) of the $\cos \theta$ distribution, for which the procedure can be taken from an existing framework that is used to study $\Lambda$ global polarization~\cite{Abelev:2007zk}. 

We can express the $\frac{dN}{d\cos \theta}$ distribution as:
\begin{eqnarray}
\frac{dN}{d\cos \theta} \sim 1 + P_{s} \cos \theta,
\label{eq:dNdCosTheta}
\end{eqnarray}
where $P_s$ quantifies the asymmetry effect, in analog to $P_{H}$ used in the hyperon global polarization study. Without the magnetic field, $P_s$ is zero, since there is an equal possibility for $\hat{s}$ to point up and down. It can be shown that
\begin{eqnarray}
\begin{aligned}
P_s  &= 3 \langle \cos \theta \rangle . \\
\end{aligned}
\label{eq:PsIntermediate}
\end{eqnarray}
In practice, the observation of $P_s$ may directly follow the steps of the global polarization measurement~\cite{Abelev:2007zk} :
\begin{eqnarray}
P_s = - \frac{8}{\pi} \frac{\big \langle \sin(\phi_s - \Psi_{\mathrm{EP}}^{(1)}) \big \rangle}{R_\mathrm{EP}^{(1)}}.
\label{eq:PsMeasurement}
\end{eqnarray}
Here $\phi_s$ is the azimuthal angle of $\hat{s}$ in the transverse plane, $\Psi_{\mathrm{EP}}^{(1)}$ is the first-order event plane angle, and $R_\mathrm{EP}^{(1)}$ is its resolution. The $P_s$ observable takes advantage of an established framework where the effects of the finite event plane resolution and the detector acceptance are well understood. The latter effect is not discussed here, but can be found in Ref.~\cite{Abelev:2007zk}.

\begin{figure}[htbp]
\centering
\makebox[1cm]{\includegraphics[width=0.45 \textwidth]{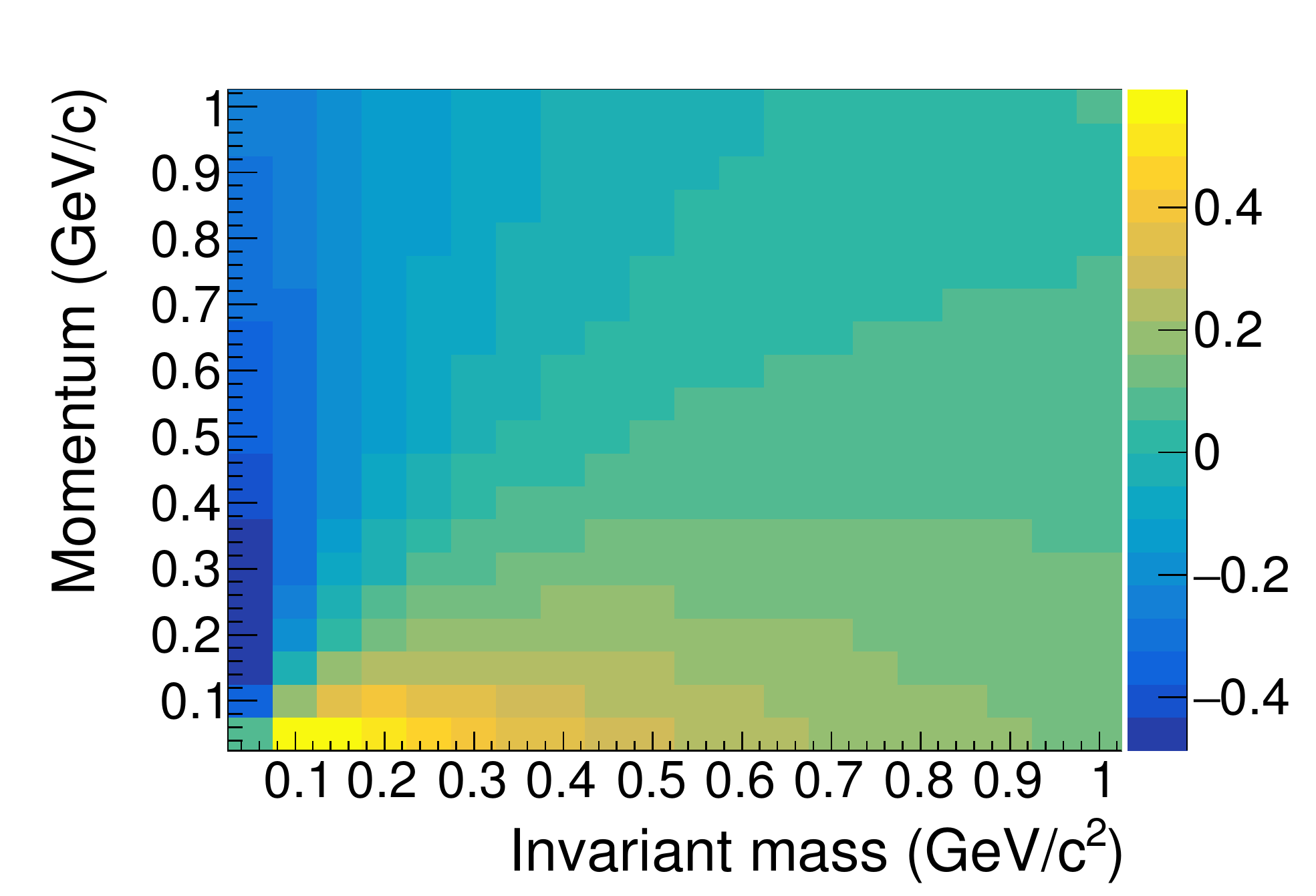}}
\caption{(Color online) $\langle \sin \Delta \alpha \rangle$ for $e^+e^-$ pairs from free-decay parents, as a function of the parents' mass and momentum. The applied magnetic field is $e B = 5 \times 10^{-3}  \mathrm{\, GeV^2} $ for a period of 1 fm/$c$. }
\label{fig:sin_alpha_pm}
\end{figure}

Note that $P_s$ depends on mass and momentum of the parent. This is more conveniently demonstrated with the quantity of $\langle \sin \Delta \alpha \rangle$. When a parent has a large mass, it tends to decay into a back-to-back $e^+e^-$ pair ($\Delta \alpha \sim \pi$), and as the magnetic field slightly decreases $\Delta \alpha$, $\langle \sin \Delta \alpha \rangle$ becomes positive. Conversely, when a parent has a high momentum, it tends to produce a near-angle $e^+e^-$ pair ($\Delta \alpha \sim 0$), and with the slight decrease of $\Delta \alpha$, $\langle \sin \Delta \alpha \rangle$ becomes negative. This is shown with a free-decay simulation in Fig.~\ref{fig:sin_alpha_pm}. On top of the two mechanisms mentioned above, when the parent's momentum or mass is large, $\gamma_{\rm RP}$ also becomes large, which counter-acts the deflection according to Eq. (\ref{eq:e_deltaAlphaChange}). This explains why in Fig.~\ref{fig:sin_alpha_pm} the largest magnitudes do not appear at infinitely large mass or momentum. Instead, they peak/dip at relatively low mass or momentum regions. Without educated selection on mass and momentum, $\langle \sin \Delta \alpha \rangle$ from positive and negative contributions can largely cancel each other, downgrading the sensitivity of the observable. Experimental search for a finite $P_s$ can use Fig.~\ref{fig:sin_alpha_pm} to tune the mass and momentum selection for an optimal signal.   
\begin{figure}[htbp]
\centering
\makebox[1cm]{\includegraphics[width=0.45 \textwidth]{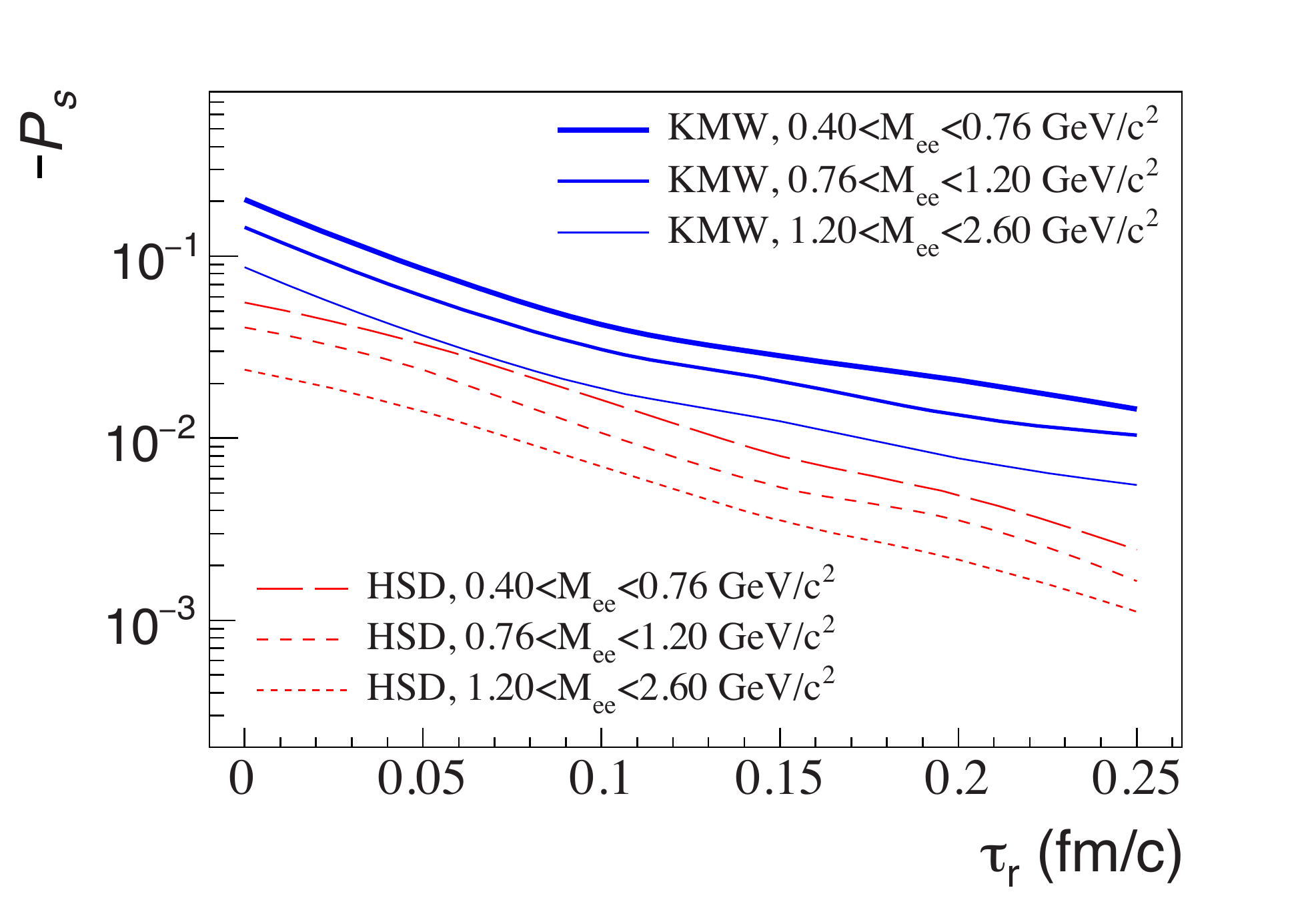}}
\caption{(Color online) ($-P_{s}$) versus relative production time ($\tau_r$) of $e^+e^-$ pairs, for two cases of the initial magnetic fields. Solid lines (KMW) are based on Fig.A.2 in Ref.~\cite{Kharzeev:2007jp}, for $b = 8 $ fm. Dashed lines (HSD) are based on Fig. 4 in Ref.~\cite{Voronyuk:2011jd}, for  $b = 10 $ fm.
 For both cases, the $P_s$ calculation is done for three divisions of $e^+e^-$ invariant mass ($M_{ee}$) according to Ref.~\cite{Adam:2018tdm}.}
\label{fig:time_dependent}
\end{figure}

The enhanced dilepton production at very low transverse momenta ($p_T$) observed at both RHIC~\cite{Adam:2018tdm} and LHC~\cite{Aaboud:2018eph} has been related to the initial electromagnetic field. Figure~\ref{fig:time_dependent} presents our $P_s$ calculation with a realistic low-$p_T$ $e^+e^-$ continuum ($p_T$ and mass spectrum taken from the STAR publication~\cite{Adam:2018tdm}). We assume that $e^+e^-$ pairs are produced at midrapidities and at a time relative to the peak time of the magnetic field, $\tau_r = \tau_\mathrm{production} - \tau_\mathrm{peakB}$. After the leptons are produced, their momentum directions are deflected by the remaining magnetic field in two cases: one calculated by Kharzeev, McLerran and Warringa (KMW)~\cite{Kharzeev:2007jp}, and the other from the approach of Hadron String Dynamics (HSD) by Voronyuk {\it et al.}~\cite{Voronyuk:2011jd}.  The resultant $P_s$ is on the order of $10^{-3}-10^{-1}$ level. In general, HSD gives lower $P_s$ values than KMW, because the retarded magnetic field in the HSD model has a lower peak magnitude and decays faster. In reality, $e^+e^-$ pairs are not produced at a single point in time, and hence one needs to use this figure to take a weighted average of $P_s$ over $\tau_r$. Note that for both the KMW- and HSD-based calculations, the field is taken at the centroid of two colliding nuclei, which is at its maximum in space. If instead one averages $B$ over space, the resultant $P_s$ will be smaller. As we have only considered the Lorentz force, the Farady's law effect will make $P_s$ smaller, and meanwhile the magnetic induction will make it larger. A full consideration requires detailed modeling, which is beyond the scope of this paper. Here we only attempt to demonstrate the feasibility of detecting $P_s$, and to provide a guidance on its order of magnitude.

As a disclaimer, a construction of the cross product of particle pairs with opposite charges has been discussed in~\cite{Kharzeev:1998kz,Finch:2001hs}. An attempt to apply $\hat{s}\cdot\hat{B}$ on $e^+e^-$ pairs to look for the magnetic field can be found in  Ref.~\cite{TanizakiThesis}, proposing a skewness study of its distribution. However, the method was applied on dileptons produced by thermal radiation, for which the signal is expected to be weak.  In our paper, we have designed a different observable ($P_s$), which can be conveniently measured with comprehensive and practical considerations. We have also offered insights on the underlying reasons for which the procedure works, as well as insights on the optimization of the signal.

\section{Signed balance function}

The magnetic field also creates a charge imbalance in the $\Delta \alpha$ space. In this section, we discuss how to quantify the magnetic field effect with slightly modified balance functions that take into account the order of the two balancing charges. We invoke two signed balance functions,% $\mathcal{B}_P$ and $\mathcal{B}_N$:

\begin{eqnarray}
\begin{aligned}
\mathcal{B}_P (\Delta \alpha) =  \frac{N_{+-}(\Delta \alpha)-N_{++}(\Delta \alpha)}{N_+},
\end{aligned}
\label{eq:Bp}
\end{eqnarray}
and
\begin{eqnarray}
\begin{aligned}
\mathcal{B}_N (\Delta \alpha) =  \frac{N_{-+}(\Delta \alpha)-N_{--}(\Delta \alpha)}{N_-}.
\end{aligned}
\label{eq:Bn}
\end{eqnarray}
Here $N_{+-}(\Delta \alpha)$ denotes the number of $e^+e^-$ pairs in a given $\Delta\alpha$ range ($\Delta \alpha =\alpha_{e^{+}}-\alpha_{e^{-}}$) in all events.
$N_{++}(\Delta \alpha)$, $N_{-+}(\Delta \alpha)$ and $N_{--}(\Delta \alpha)$ are defined in a similar way, except that $\Delta \alpha$ becomes $(\alpha_{e^{+}}-\alpha_{e^{+}})$, $(\alpha_{e^{-}}-\alpha_{e^{+}})$ and $(\alpha_{e^{-}}-\alpha_{e^{-}})$, respectively. 
Note that unlike the uniquely defined $\Delta \alpha$ as in Sec.~\ref{sec:distortion} and ~\ref{sec:SDotB}, now different combinations of the subscripts $a$ and $b$ in $N_{ab}$ correspond to different constructions of $\Delta \alpha$.
$N_{+(-)}$ is the number of positrons (electrons) integrated over all events. 

\begin{figure}[htbp]
\centering
\resizebox{8cm}{!}{\includegraphics{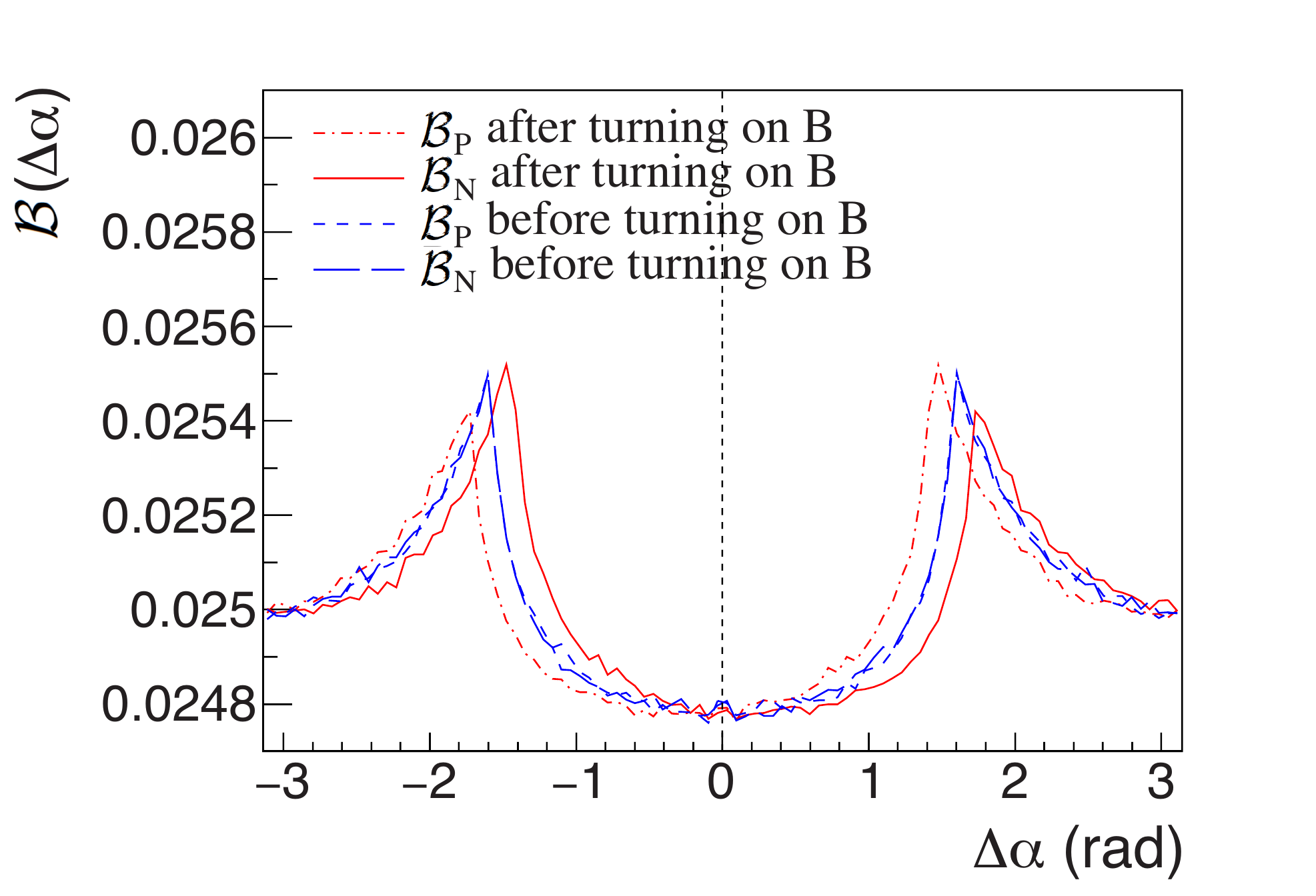}}
%\makebox[1cm]{\includegraphics[width=0.45 \textwidth]{fig7_bfunc-eps-converted-to.pdf}}
\caption{(Color online) $\mathcal{B}_P$ and $\mathcal{B}_N$ calculated for $e^+e^-$ pairs from virtual-particle decays, with and without the magnetic field. The magnetic field and its acting time, mass and momentum of the virtual particle are the same as those in Fig.~\ref{fig:compare_rand_sig}. }
\label{fig:bfunc}
\end{figure}

In general, the balance function reflects the absolute separation of particles in phase space~\cite{Bass:2000az,Adams:2003kg}. For example, the balance function in pseudorapidity, $\mathcal{B}(\Delta \eta)$, spans the absolute difference in pseudorapidity between two balancing particles, $\Delta \eta = |\eta_a - \eta_b|$. Here we consider the signed difference, instead of the absolute difference, in the $\alpha$ space. We also investigate $\mathcal{B}_P$ and $\mathcal{B}_N$ separately, instead of the average of the two. For completeness, the standard balance function can be recovered as: 
\begin{eqnarray}
\mathcal{B}(| \Delta \alpha |) = \frac{1}{2} [\mathcal{B}_P(|\Delta \alpha |) + \mathcal{B}_N( |\Delta \alpha |)]. 
\label{eq:BalanceFcn}
\end{eqnarray}

\begin{figure}[htbp]
\centering
\resizebox{8cm}{!}{\includegraphics{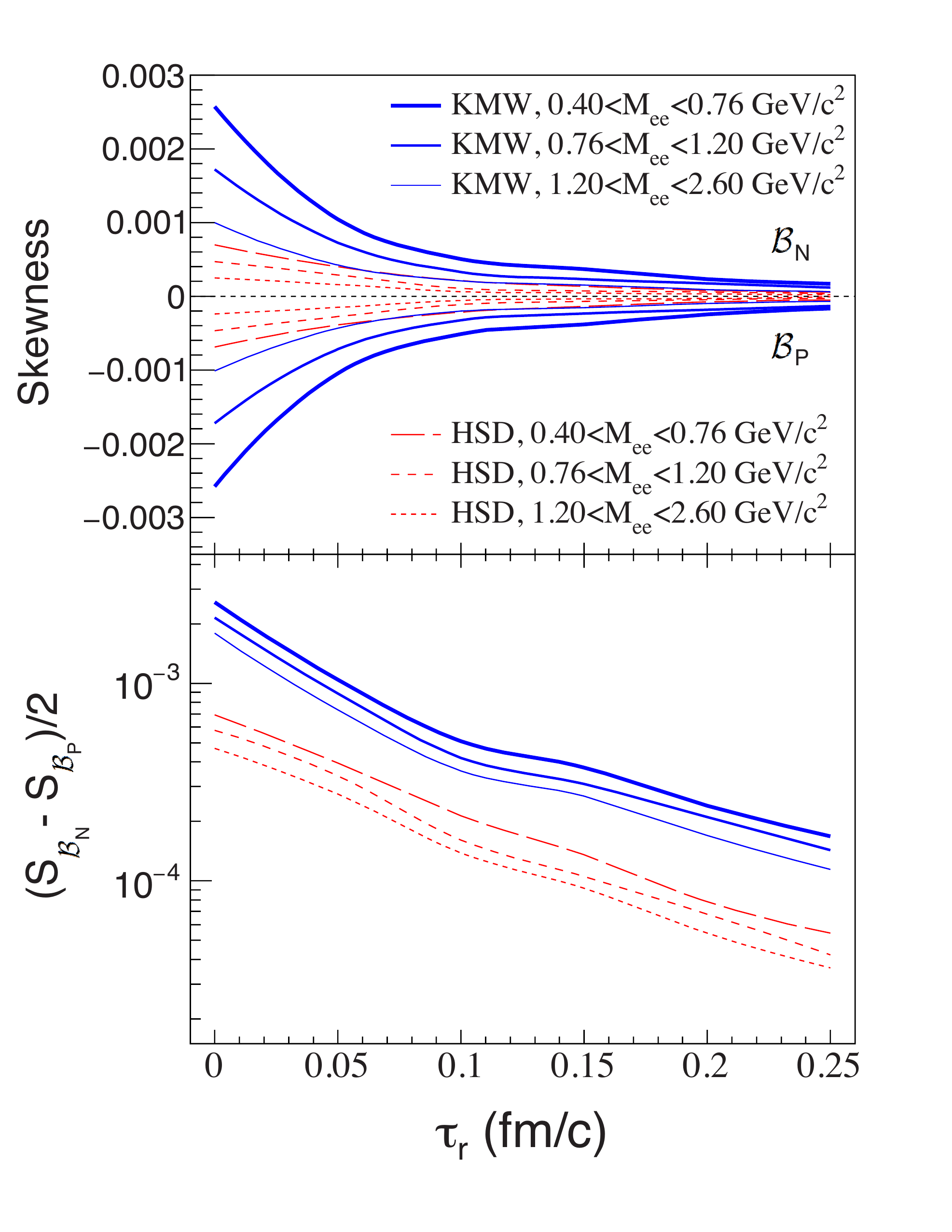}}
%\makebox[1cm]{\includegraphics[width=0.45 \textwidth]{fig8_bfunc_time-eps-converted-to.pdf}}
\caption{(Color online) Upper panel: skewness of $\mathcal{B}_{P}$ and $\mathcal{B}_{N}$ versus relative production time ($\tau_r$) of $e^+e^-$ pairs, for two cases of the initial magnetic field. Solid lines (KMW) are based on Fig.A.2 in Ref~\cite{Kharzeev:2007jp}, for $b = 8 $ fm. Dashed lines (HSD) are based on Fig. 4 in Ref~\cite{Voronyuk:2011jd}, for  $b = 10 $ fm. For both cases, the calculation is done for three divisions of $e^+e^-$ invariant mass ($M_{ee}$) according to Ref.~\cite{Adam:2018tdm}. Lower panel: the same calculation, but presented with normalized skewness difference, $\frac{S_{\mathcal{B}_N} - S_{\mathcal{B}_P}}{2}$, in log scale for clear view of small values. }
\label{fig:bfunc_time}
\end{figure}

These two signed balance functions, $\mathcal{B}_P$ and $\mathcal{B}_N$, are subject to the effect of the magnetic field. Figure~\ref{fig:bfunc} shows our simulations of $\mathcal{B}_P$ and $\mathcal{B}_N$ with a magnetic field acting upon $e^+e^-$ pairs from virtual-particle decays. Here the magnetic field, mass and momentum of the virtual particle are the same as those in Fig.~\ref{fig:compare_rand_sig}. Without the magnetic field, the two functions coincide. Conversely, when the magnetic field is on, for the same reason mentioned in Sec.~\ref{sec:distortion}, the two functions are skewed away from each other, with $\mathcal{B}_P$ ($\mathcal{B}_N$) shifting towards the negative (positive) $\Delta \alpha$ direction. In fact, the magnetic field will also modify the width of the standard balance function, $\mathcal{B}(| \Delta \alpha |)$. However, such information is not easily obtainable, since it is elusive to define a good reference for the case without the magnetic field. On the other hand, with the modified balance function we suggest, the reference is clear for the case without the magnetic field: $\mathcal{B}_N$ and $\mathcal{B}_P$ have to be identical, and both are symmetric w.r.t $\Delta \alpha = 0$, according to simple symmetry arguments.

In Fig.~\ref{fig:bfunc_time},  we conduct a similar exercise as in Fig.~\ref{fig:time_dependent}, except that here we calculate the skewness of $\mathcal{B}_P$ and $\mathcal{B}_N$ instead of $P_s$. The skewness values for $\mathcal{B}_P$ and $\mathcal{B}_N$ bear opposite signs, as expected. The values are on the order of $10^{-4} - 10^{-3}$ level. This information serves as a reference to gauge the experimental measurements of the magnetic field using these signed balance functions.

As mentioned in the previous section, the event plane resolution can be conveniently corrected for the $P_s$ measurements. For the approach presented in this section, the effect due to the event plane resolution has to be studied through simulation. On the other hand, the extraction of the skewness of $\mathcal{B}_P$ and $\mathcal{B}_N$ does not require the reconstruction of mixed events for background subtraction (that is needed for the $P_s$ calculation), because the balance function by design has already subtracted the same-sign pairs, serving the purpose of background subtraction.

\section{Sensitivity study}

\begin{figure}[htbp]
\centering
\resizebox{8cm}{!}{\includegraphics{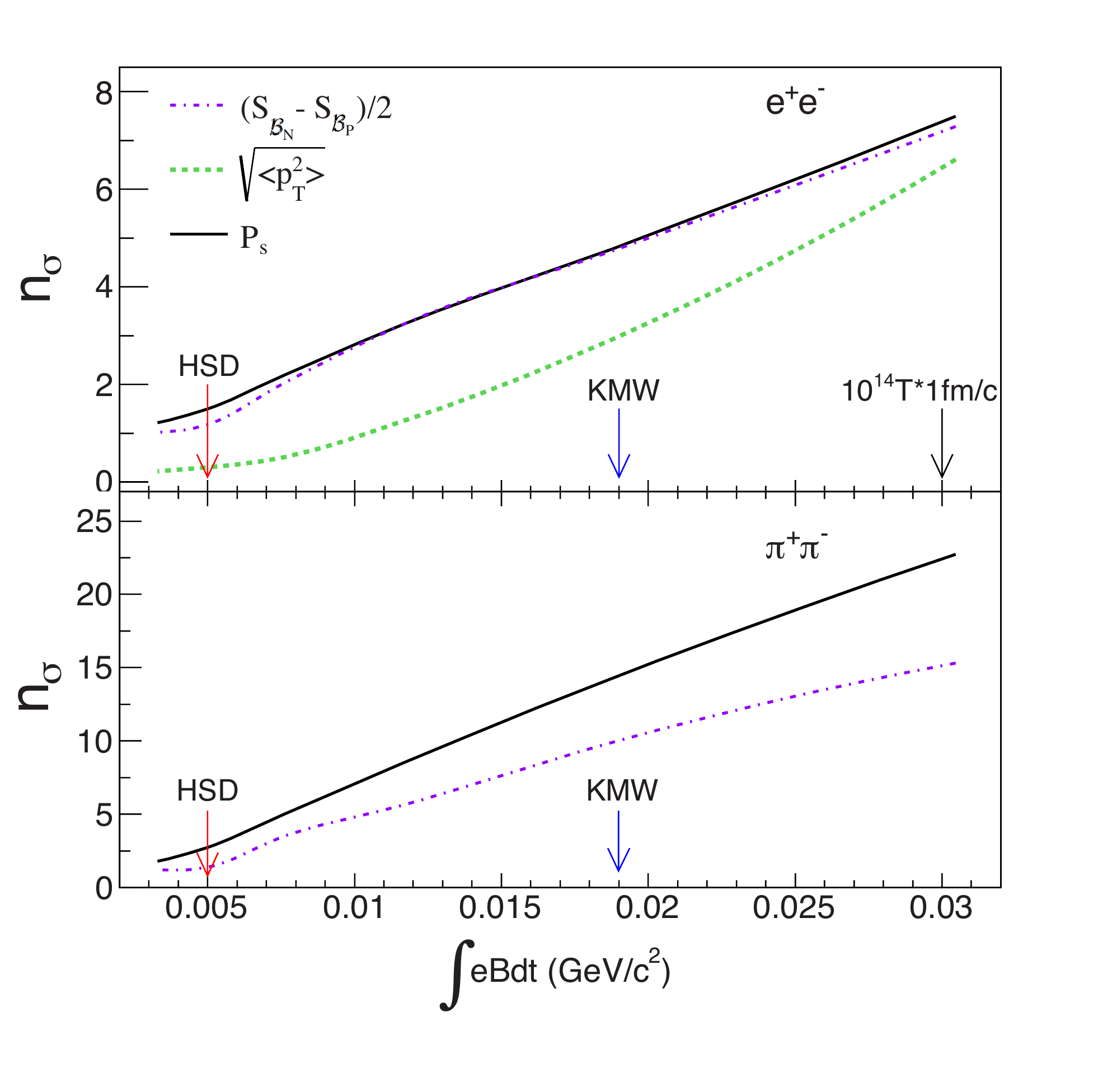}}
%\makebox[1cm]{\includegraphics[width=0.45 \textwidth]{fig7_bfunc-eps-converted-to.pdf}}
\caption{The sensitivity for $e^+e^-$ (top panel) and $\pi^+\pi^-$ (bottom panel) pairs, based on studies with simulated pairs. See text for details. The invariant mass spectrum for signal pair of $\pi^+\pi^-$ is taken from ~\cite{Zhao:2017nfq}.}
\label{fig:sensitivity}
\end{figure}

In order to understand the sensitivity of different observables, we simulate $e^+e^-$ pairs with the same initial $p_T^2$ spectra (before applying the EM effect) as used to study the $p_T$ broadening in Ref.~\cite{Adam:2018tdm}, which is based on the calculations by Zha {\it et al.}~\cite{Zha:2018ywo}. As shown with the dashed line in the top panel of Fig.~\ref{fig:sensitivity}, we tune the statistics to achieve the $6\sigma$ significance for the change of $\sqrt{\langle p_T^2 \rangle}$ (the same as stated  in Ref.~\cite{Adam:2018tdm}), with a constant magnetic field of $10^{14}$ T that lasts 1 fm/$c$, within the invariant mass range of $0.4-0.76$ GeV/$c^2$. The mass window, applied magnetic field and the duration it lasts are also the same as used in ~\cite{Adam:2018tdm}.  With the same data sample, the significance values ($n_{\sigma}$) for the $P_s$ and for the skewness of the balance function, $(S_{B_N}-S_{B_P})/2$, are also shown as functions of $\int e B dt$ with a solid line and a dot-dashed line, respectively. The two approaches proposed in this paper clearly outperform the approach based on $\sqrt{\langle p_T^2 \rangle}$. This is expected because the parent $p_T$ is indirectly affected by the magnetic field via the distortion of the relative angle ($\Delta \alpha$) between the two daughters. Conversely, the $P_s$ and the skewness of the balance function directly probe the change in $\Delta \alpha$. We also mark the $\int e B dt$ values for the two cases of the magnetic field calculations (HSD and KMW), both of which are time dependent, instead of being constant. When more realistic magnetic fields are employed, the significance is largely reduced, compared with the case of the constant magnetic field. 

The bottom panel of Fig.~\ref{fig:sensitivity} shows the sensitivity for $\pi^+\pi^-$ pairs with a simplified, conservative assumption that only one $\rho$ meson per event is detected through the $\pi^+\pi^-$ channel, although $dN/dy$ for the $\rho$ production is $\sim5.4$~\cite{Adams:2003cc}.   Pions are heavier than electrons, and pion pairs are produced later than $e^+e^-$ pairs, but they outnumber $e^+e^-$ pairs by a large amount. The absolute significance depends on many factors, such as the production time of pion/electron pairs, the strength of the magnetic field, its sustainability in the QGP medium, and so on. The intention here is not to give a definite guidance on the significance, but to point out that $\pi^+\pi^-$ pairs are also worth investigating owing to their relatively large statistics.      

\section{Summary}

We have investigated the imprint left by the initial magnetic field on pairs of oppositely charged particles in relativistic heavy-ion collisions. The underlying mechanism is the distortion of the relative angle between positively- and negatively-charged particles inside a pair. We adopt two observables to detect this effect: one based on the framework for measuring the hyperon global polarization, and the other based on the balance function with slight modifications. We have estimated the magnitude ranges and studied sensitivities for the two observables in the case of the Lorentz force. The knowledge documented in this paper will facilitate the experimental efforts to quantify the strong magnetic field in high-energy nuclear collisions. 

\section*{Acknowledgements}
We'd like to thank J. Liao, Q. Y. Shou, S. Yang and Z. Xu for fruitful discussions.  Y.J. Ye and Y.G. Ma are supported in part by the National Natural Science Foundation of China under Contracts No. 11890710, No. 11890714, and No. 11421505, the Key Research Program of Frontier Sciences of the CAS under Grant No. QYZDJ-SSW-SLH002, and the Key Re- 364 search Program of the CAS under Grant No. XDPB09. A.H. Tang is supported by the US Department of Energy under Grants No. DE-AC02-98CH10886 and No. DE-FG02-89ER40531. G. Wang is supported by the US Department of Energy under Grant No. DE-FG02-88ER40424.

{}
\end{document}